\documentclass[USenglish,twocolumn]{article}

\ifx\directlua\undefined\ifx\XeTeXcharclass\undefined
\usepackage[utf8]{inputenc}                           
\else\RequirePackage[no-math]{fontspec}[2017/03/31]\fi 
\else\RequirePackage[no-math]{fontspec}[2017/03/31]\fi 
\usepackage[sort&compress,square,numbers]{natbib}
\usepackage[big,online]{dgruyter}
\usepackage{color}
\usepackage{bm}
\usepackage{amsmath,amsthm,amssymb,amsfonts}
\usepackage{hyperref}
\hypersetup{
	colorlinks=true,
	linkcolor=blue,
	filecolor=blue,
	urlcolor=blue,
	citecolor=blue,
}
\theoremstyle{dgthm}

\theoremstyle{dgdef}

\newcommand{\bk}{{\bm k}}
\newcommand{\bn}{{\bm n}}
\newcommand{\br}{{\bm r}}
\newcommand{\bv}{{\bm v}}

\begin{document}
	
	\articletype{Research Article}
	
	\author[1]{Shuang Shen}
	\author*[1]{Yiqi Zhang} 
	\author[2]{Yaroslav V. Kartashov}
	\author[1]{Yongdong Li}
	\author[3]{Vladimir V. Konotop}
	\affil[1]{Key Laboratory for Physical Electronics and Devices, Ministry of Education, School of Electronic Science and Engineering, Xi'an Jiaotong University, Xi'an 710049, China,
		E-mail: shenshuang2023@126.com (S. Shen), zhangyiqi@xjtu.edu.cn (Y. Zhang), leyond@xjtu.edu.cn (Y. Li). 
		https://orcid.org/0000-0001-9066-5580 (S. Shen).
		https://orcid.org/0000-0002-5715-2182 (Y. Zhang).
		https://orcid.org/0000-0001-8756-0438 (Y. Li).    }
	\affil[2]{Institute of Spectroscopy, Russian Academy of Sciences, Troitsk, Moscow, 108840, Russia, Email: yaroslav.kartashov@icfo.eu. https://orcid.org/0000-0001-8692-982X}
	\affil[3]{Departamento de F\'{\i}sica and Centro de F\'{\i}sica Te\'{o}rica e Computacional, Faculdade de Ci\^{e}ncias, Universidade de Lisboa, Campo Grande, Ed. C8, Lisboa 1749-016, Portugal, Email: vvkonotop@fc.ul.pt. https://orcid.org/0000-0002-1398-3910}
	
	\title{Two-dimensional flat-band solitons in  superhoneycomb lattices}
	\runningtitle{Flat-band solitons in  superhoneycomb lattices}
	\abstract{Flat-band periodic materials are characterized by a linear spectrum containing at least one band where the propagation constant remains nearly constant irrespective of the Bloch momentum across the Brillouin zone. These materials provide a unique platform for investigating phenomena related to light localization. Meantime, the interaction between flat-band physics and nonlinearity in continuous systems remains largely unexplored, particularly in continuous systems where the band flatness deviates slightly from zero, in contrast to simplified discrete systems with exactly flat bands. Here, we use a continuous superhoneycomb lattice featuring a flat band in its spectrum to theoretically and numerically introduce a range of stable flat-band solitons. These solutions encompass fundamental, dipole, multi-peak, and even vortex solitons. Numerical analysis demonstrates that these solitons are stable in a broad range of powers. They do not bifurcate from the flat band and can be analyzed using Wannier function expansion leading to their designation as \textit{Wannier solitons}. These solitons showcase novel possibilities for light localization and transmission within nonlinear flat-band systems.}
	\journalname{Nanophotonics}
	\journalyear{2024}
	\journalvolume{aop}
	
	\maketitle
	
	\section{Introduction}
	
	Existence of a flat band or of several flat bands in the spectrum of a linear Hamiltonian significantly changes the properties of a system. This is well understood theoretically and verified experimentally in diverse areas of physics ranging from solid-state physics, where flat bands appear in bi-layer graphene, to atomic and optical mono-layered systems, where flat bands arise in spectra of some lattice potentials~\cite{leykam.aplp.3.070901.2018, leykam.aipx.3.1473052.2018, tang.nano.9.1161.2020, vicencio.aipx.6.1878057.2021, chekelsky.nrm.2024}. The majority of studies of optical flat-band systems was concentrated on discrete settings, which represent tight-binding limits of the respective continuous models with periodic refractive index landscapes. From the theoretical point of view discrete lattices benefit from the existence of {\em exact} flat bands and from the possibility of algorithmic design of structures that feature such bands~\cite{morales.pra.94.043831.2016}. From the experimental point of view, discrete models adequately describe light propagation dynamics in arrays of sufficiently deep waveguides~\cite{vicencio.prl.114.245503.2015,mukherjee.prl.114.245504.2015,zhang.aop.382.160.2017,hanafi.apl.7.111301.2022,xia.ol.41.1435.2016,zong.oe.24.8877.2016,travkin.apl.111.011104.2017,xia.prl.121.263902.2018,ma.prl.124.183901.2020,yan.aom.8.1902174.2020,song.lpr.17.2200315.2023} representing also one of the most powerful platforms for the exploration of self-action of light in periodic environment, but produce considerable deviations from observed dynamics in lattices with smooth refractive index landscapes or in arrays with relatively small refractive index contrast. Very recently, photonic structures with multiple flat bands on the basis of microwave resonators were reported~\cite{yang.nc.15.1484.2024}.
	
	Exactly flat bands in discrete systems have opened opportunities for construction of new types of self-sustained states. Compactons and solitons were shown to exist in both conservative and dissipative discrete arrays~\cite{vicencio.pra.87.061803.2013}. Two-dimensional thresholdless solitons bifurcating from linear compact modes have been found in kagome arrays~\cite{yulin.ol.38.4880.2013}. Flat-band solitons and nonlinear localized flat-band modes were also reported in sawtooth-like lattices~\cite{johansson.pre.92.032912.2015}, diamond chain lattices~\cite{gligoric.prb.94.144302.2016,zegadlo.pre.96.012204.2017}, Stub lattices~\cite{baboux.prl.116.066402.2016,goblot.prl.123.113901.2019}, Lieb lattices~\cite{belicev.pra.96.063838.2017,lazarides.prb.96.054305.2017,real.pra.98.053845.2018, ali.pra.103.013517.2021}, and octagonal-diamond lattices~\cite{stojanovic.pra.102.023532.2020}. A summary of recent theoretical and experimental advances in the area of multidimensional localized structures in optical media, including the formation of different types of lattice solitons in discrete and continuous physical models can be found in \cite{mihalache.rrp.76.402.2024}.
	
	However, in genuinely continuous systems, whose accurate description requires going beyond tight-binding approximations, exactly flat bands do not exist. Instead, one can realize continuous systems with {\em nearly flat} bands, whose width is sufficiently small, but not exactly zero. We emphasize that here the flatness of the band is implied in all directions of the reduced Brillouin zone (BZ). This opens the important and so far unaddressed question about the diversity, existence, and bifurcations of self-sustained nonlinear states in such two-dimensional (2D) continuous waveguiding systems. 
	
	Some preliminary conclusions that the behaviour of nonlinear states in such systems can be very unusual and can sharply contrast with behaviour of discrete systems, can be drawn on the basis of recent studies of Wannier solitons in 1D flat-band system describing spin-orbit coupled Bose-Einstein condensates~\cite{wang.pra.108.013307.2023} and studies of light propagation in photonic moir\'e lattices~\cite{wang.nature.577.42.2020}, where nearly all higher bands of the optical potential are flat, allowing excitation of 2D solitons practically with zero power threshold~\cite{fu.np.14.663.2020}.
	
	In addition to obvious limitations of the tight-binding (discrete) models that neglect other (non-flat) bands which are necessarily excited in nonlinear systems, there is also a conceptual difference consisting in certain ambiguity of such models. Indeed, deriving a discrete model from a continuous one requires definition of a proper basis. Even restricting the consideration to the most natural choice of a basis of Wannier functions (WFs), the latter are not uniquely defined, what is particularly relevant for two- and three-dimensional systems~\cite{marzari.rmp.84.1419.2012}. The coupling coefficients of the respective one-band discrete approximation are determined by the hopping of the chosen basis functions, and thus are not uniquely defined for a given continuous model, as well. On the other hand, different continuous models may result in the same equation of the tight-binding approximation. This does not allow one to make one-to-one correspondence between the original nonlinear continuous model and its tight-binding approximation, what makes direct study of the continuous models particularly relevant.
	
	In this work we use continuous superhoneycomb lattices that possesses a nearly flat band in their linear spectrum, to study the emergence and stability properties of solitons of very different types. Our choice of the super-honeycomb lattice is dictated by its proprieties which are known due to previous studies~\cite{zhong.adp.529.1600258.2017, lan.prb.85.155451.2012, yan.aom.8.1902174.2020} allowing us to straightforwardly find the parameter enabling a nearly flat band. 
	By changing the depths of some waveguides it is possible to tune the band structure such that forbidden gap appears either above or below the flat band, as shown in Fig.~\ref{fig1} in the following text. This allows to obtain flatband solitons in this continuous system both in the medium with focusing (above the flat band) and defocusing (below the flat band) nonlinearity. In addition the degree of the band flatness that one can achieve for this type of the lattice exceeds band flatness for more traditional Lieb and kagome lattices with the same depth and waveguide spacing.
	We report families of flat-band soliton and vortex-soliton solutions and show that in continuous lattices, unlike in their discrete counterparts, they are not thresholdless anymore. The  thresholds for solitons of very different types, such as fundamental and vortical ones, may be very small and comparable.
	
	\section{Flat bands in superhoneycomb lattices}
	
\begin{figure}[ht]
		\centering
		\includegraphics[width=\columnwidth]{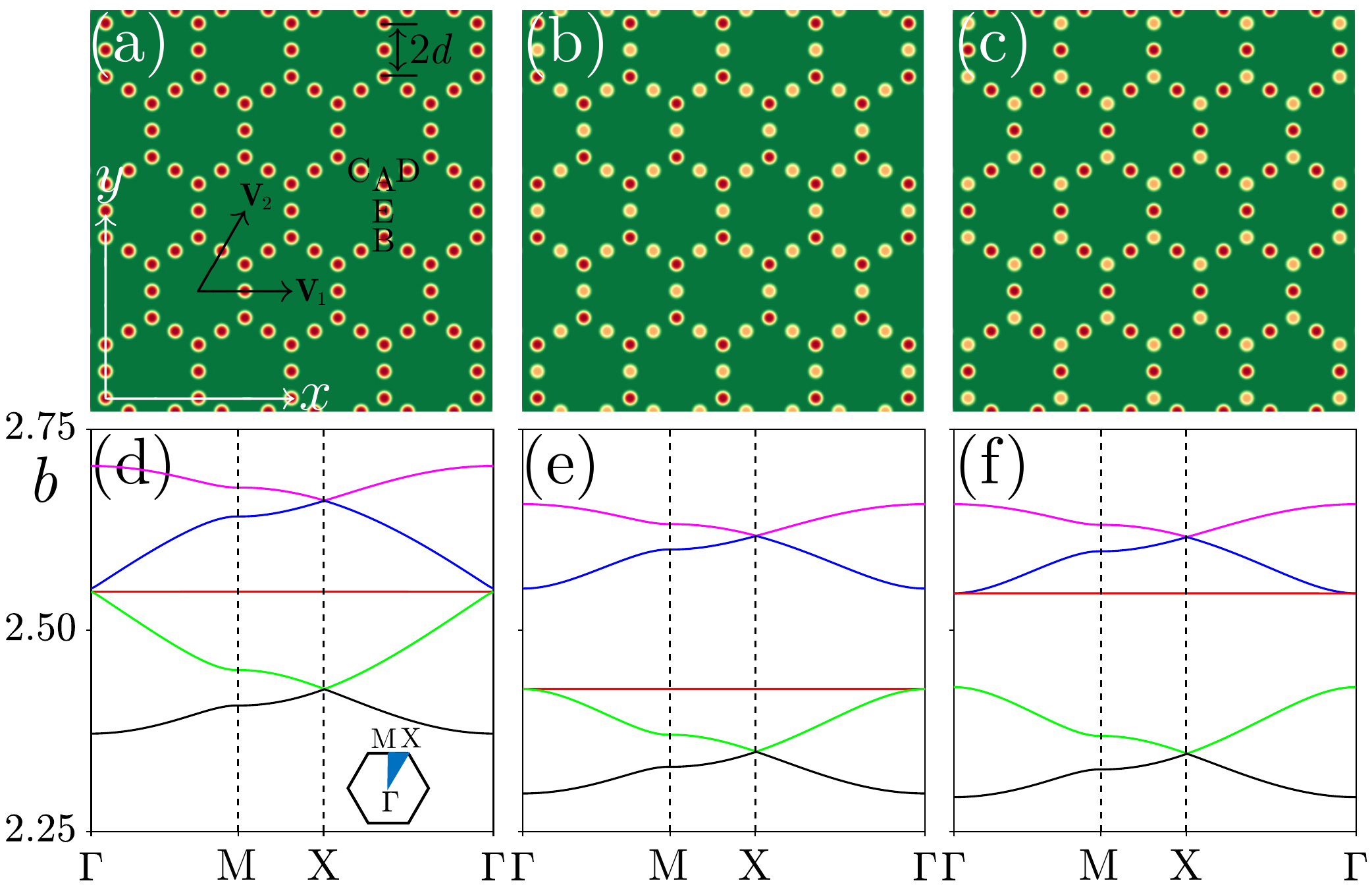}
		\caption{(a) Superhoneycomb lattice, where all waveguides have the same depth ${p=8.0}$. The basis vectors of the Bravais lattice $\bv_{1,2}$, the distance $d$ between two sites, and labels of sites within primitive lattice cell are shown. (b) Superhoneycomb lattice with depths of sites A and B ${p_{\bn_A}=p_{\bn_B}=8.0}$ and depths of sites C, D, E ${p_{\bn_C}=p_{\bn_D}=p_{\bn_E} =7.8}$. We use the notation $\bn_A$ for positions of A sites, and similar notations $\bn_D$,...,$\bn_E$ for positions of other sites. (c) Superhoneycomb lattice with ${p_{\bn_A}=p_{\bn_B}=7.8}$ and ${p_{\bn_C}=p_{\bn_D}=p_{\bn_E} = 8.0}$. (d-f) Band structures of lattices depicted in panels (a-c), respectively. The flat band present in the spectrum is indicated by red line in (d) and is preserved even when the gap appears in (e,f).}
		\label{fig1}
	\end{figure}

	The dimensionless envelope $\Psi$ of a paraxial light beam propagating along the $z-$direction in a 2D optical lattice, described by the function $\mathcal{R}(\br)$, where  ${\br=(x,y)}$, is governed by the nonlinear Schr\"odinger  (NLS) equation
	\begin{align}
		\label{NLS}
		i \frac{\partial \Psi}{\partial z}=H\Psi - \sigma|\Psi|^2\Psi,
	\end{align}
	with the linear Hamiltonian 
	\begin{align}
		\label{H}
		H=   -\frac{1}{2} \nabla^2 
		-\mathcal{R}(\br)
	\end{align}
	where ${\nabla=(\partial/\partial x,\partial/\partial y)}$ and ${\sigma=\pm1}$ corresponds to the Kerr medium either with a focusing ($\sigma=+1$) or defocusing ($\sigma=-1$) nonlinearity. In this work we focus on stationary states of the problem described by  Equations~(\ref{NLS}) and (\ref{H}). Such states have the form ${\Psi=\psi(\br)\exp(ibz)}$, where $b$ is the propagation constant, and the function $\psi(\br)$ describing the transverse profile of the mode, solves the stationary nonlinear problem
	\begin{align}
		\label{NLS-stat}
		-b\psi=H\psi - \sigma |\psi|^2 \psi.
	\end{align}
	While the theory reported below is not restricted to any specific type of optical potential (except the requirement to possess a nearly flat band), in all numerical simulations we employ the superhoneycomb lattice described by 
	\begin{align}
		\label{R}
		\mathcal{R}(\br)=\sum_{\bn} p_{\bn} \exp\left(-\frac{(\br-\bv_{\bn})^4}{w^4}\right).
	\end{align}
	Here ${\bn=[n_1,n_2]}$ with integers $n_{1,2}$ defines the positions of the lattice ``sites'': ${\bv_{\bn} = n_1 \bv_1 + n_2 \bv_2}$, where ${\bv_{1,2}}$ are the basis vectors of the Bravais lattice, $p_{\bn}$ stands for the depth of each waveguide, and $w$ is the waveguide width. The profile of the lattice for the particular case, when all $p_{\bn}$ are equal, $p_{\bn}=p$, is shown in Figure~\ref{fig1}(a).
	
	The spectrum of the underlying linear Hamiltonian in Equation~(\ref{H}), is determined by the 2D linear problem
	\begin{align}
		\label{linear}
		H\varphi_{\nu \bk}=-b_\nu (\bk)\varphi_{\nu \bk}
	\end{align}
	for 2D linear Bloch modes ${\varphi_{\nu \bk}(\br)=e^{i\bk\cdot\br} u_{\nu \bk}(\br)}$, where ${u_{\nu \bk}(\br)=u_{\nu k}(\br+\bv_\bn)}$, $\bv_\bn$ is a lattice vector, $\nu$ is the index of the band, and ${\bk}$ is the Bloch vector in the first BZ. In Figure~\ref{fig1}(d) we show the spectrum of the problem corresponding to the lattice in Figure~\ref{fig1}(a) (it was obtained using plane-wave expansion method and details on this method can be find in the \textbf{Methods} section), where a flat band is present, as indicated by the red line. The width of the flat band with index $\nu$ given by 
	\[
	{\Delta_\nu=\max_{\bk\in \text{BZ}}b_\nu (\bk)-\min_{\bk\in \text{BZ}}b_\nu (\bk)}
	\]
	is not exactly zero and can be used as a parameter characterizing band flatness. The ideally flat band would have ${\Delta_\nu=0}$, while for the band shown in Figure~\ref{fig1}(d) one has ${\Delta_\nu \approx 2.77\times10^{-5}}$. Notice that $\Delta_\nu$ increases with decrease of the lattice depth $p$. We stress that we consider here shallow periodic photonic structures defined by small refractive index modulations. For example, the dimensionless lattice depth of ${p=k^2r_0^2\delta n/n \sim 8}$ used here corresponds to the refractive index contrast ${\delta n \sim 8 \times 10^{-4}}$, where ${k=2\pi n/\lambda}$ is the wavenumber at the wavelength ${\lambda=800~\textrm{nm}}$, ${n=1.45}$ is the unperturbed refractive index of the material (fused silica, for example), ${r_0=10~\mu \textrm{m}}$ is the characteristic transverse scale to which coordinates $x,y$ are normalized. Thus the Equation~(\ref{NLS}) is ideally suited for accurate description of the paraxial light propagation in such structures~\cite{rechtsman.nature.496.196.2013, kirsch.np.17.995.2021, ren.light.12.194.2023, arkhipova.sb.68.2017.2023, zhong.ap.3.056001.2021, zhang.elight.3.5.2023, wang.nature.577.42.2020, wang.np.18.224.2024}. 
	
	The existence of gap solitons hinges on the requirement that the flat band is isolated from the rest of the spectrum by a gap, positioned either below or above the band. Examples of such situations are known~\cite{lan.prb.85.155451.2012, zhong.adp.529.1600258.2017, tang.rrp.74.504.2022}. To make possible the existence of gap solitons in our case, we notice that each primitive cell of the honeycomb lattice has five waveguides, denoted by letters A to D in Figure~\ref{fig1}(a), and judiciously detune the depths of two groups of waveguides by making $p_{\bn}$ different. Thus, when the waveguides A and B have larger depths than other waveguides, as shown in Figure~\ref{fig1}(b), a gap appears above the flat band, while the width of this band practically does not increase and is equal to ${\Delta_\nu \approx 3.54\times10^{-5}}$ [see red line in Figure~\ref{fig1}(e)]. Meantime no gap opens below this flat band that touches the next band $\nu+1$ at the center of the BZ. On the other hand, if the depths of waveguides A and B are reduced in comparison with depths of other three waveguides in the unit cell [Figure~\ref{fig1}(c)], then a relatively wide gap opens below the flat band, that in this case keeps its width at ${\Delta_\nu \approx 2.94\times10^{-5}}$, as shown by the red line in Figure~\ref{fig1}(f).  
	
	\section{Two-dimensional Wannier solitons and vortex solitons: numerical study}
	\label{sec:Solit-Vortex}
	
	We are interested in properties of gap solitons, i.e. square integrable solutions, whose propagation constants $b$ belong to a gap adjacent to the flat band. We consider here only static solitons that do not move across the lattice, i.e. their spatial group velocity is zero, while field modulus distribution does not change with propagation distance $z$. Families of such solutions can be parameterized by the total power ${P(b)=\int_{\br\in\mathbb{R}^2} |\Psi|^2d\br}$ carried by the mode. Let us introduce the notation ${b_{\rm co}=\max_{\bk\in \text{BZ}}b_\nu (\bk)}$ (${b_{\rm co}=\min_{\bk\in \text{BZ}}b_\nu (\bk)}$) in the case of a gap above (below) the flat band corresponding to the case illustrated in Figure~\ref{fig1}(e) (Figure~\ref{fig1}(d)). In 2D lattices with non-flat bands, 2D solitons carrying a finite total power cannot bifurcate from linear Bloch states in a sense that $P(b)$ does not tend to zero in the limit ${b\to b_\textrm{co}}$~\cite{ilan.siam.8.1055.2010} (see also numerical studies in~\cite{shi.pre.75.056602.2007,shi.pra.78.063812.2008}), and hence, there exists a minimal threshold power $P_{\rm th}$ which is achieved at a certain ${b=b_{\rm th}}$: ${P(b)\geq P_{\rm th}=P(b_{\rm th})}$ (such constraint does not exist in discrete models). The amplitude of a nonlinear solution still vanishes at ${b\to b_\textrm{co}}$. However, when a solution forms in a gap above or below the flat band, one observes a different picture, and the dependence $P(b)$ acquires qualitatively new properties. We illustrate this first using direct numerical calculation of the families of the flat-band solitons. Such solutions can be found using standard Newton method~\cite{yang.book.2010}. To check their stability, we introduce a small-scale perturbation into soliton profiles and model their long-distance propagation according to Equation~(\ref{NLS}). If the profile of soliton remains unchanged upon propagation, we conclude that it is stable, otherwise it is considered unstable. We also would like to stress that the presence of optical lattice potential suppresses critical collapse that otherwise could occur for solitons in uniform focusing nonlinear medium.
	
	\begin{figure}[t]
		\centering
		\includegraphics[width=\columnwidth]{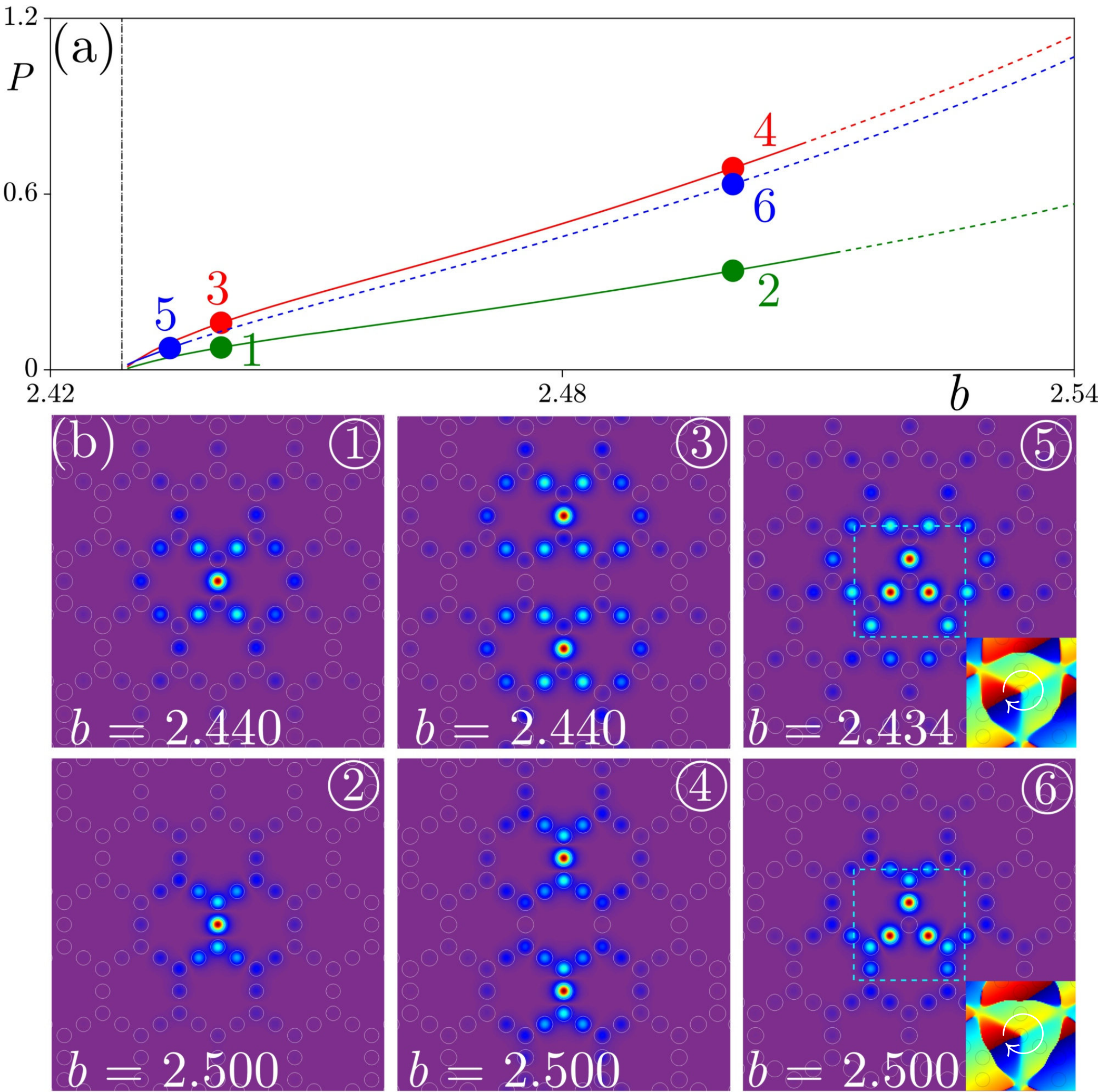}
		\caption{(a) Flat-band soliton families for focusing nonlinearity. The green, red and blue curves show families of fundamental, dipole, and vortex solitons, respectively. Solid and dashed curves represent stable and unstable solitons. The vertical dashed line at ${b=2.4283}$ indicates the  location of the flat band, whose width cannot be discerned on the scale of the figure.
			The families are terminated at at the propagation constant ${b=2.429}$. 
			(b) Field modulus distributions $|\psi|$ in selected flat-band solitons numbered 1-6. Panels in (b) are shown within ${-15\le x,y\le15}$ window. }
		\label{fig2}
	\end{figure}
	
	\subsection{Focusing case}
	\label{sec:focus}
	
	We start with the focusing case ${\sigma=+1}$ when gap solitons exist above the flat band in the lattice depicted in Figure \ref{fig1}(b). Figure~\ref{fig2}(a) shows three different families of solitons with the vertical dashed line representing the location of the flat band. The lowest (green) family corresponds to the fundamental solitons. In a flat-band system such solitons remain surprisingly well localized even very close to the band, i.e., even at relatively small values of detuning ${\delta b = b-b_{\rm co}}$ from the band [see soliton 1 shown in Figure~\ref{fig2}(b)]. When $\delta b$ increases, the localization of soliton moderately increases too [see soliton 2 shown in Figure~\ref{fig2}(b)]. Fundamental solitons are stable in the largest part of the gap, but become unstable close to its upper edge. One can use such fundamental solitons for construction of a high-order solitons that can be viewed as coupled fundamental states placed one next to the other. For example, we can use two fundamental solitons $\psi_1$ and $\psi_2$ to construct an initial guess via ${\psi_1 (y+\delta y)-\psi_2(y-\delta y)}$ with $\delta y$ being a shift in $y$. Using this guess one can obtain dipole soliton via Newton iterations [see soliton 3 in Figure~\ref{fig2}(b)]. Dipole solitons also demonstrate stability in the largest part of the gap [see red solid line in Figure~\ref{fig2}(a)], becoming unstable only for large enough values of detuning $\delta b$ [see red dashed line in Figure~\ref{fig2}(a)]. We also found a family of vortex solitons shown by the blue line in Figure~\ref{fig2}(a). The stability region of vortex solitons, however, is much smaller as compared to that for fundamental or dipole solitons. Representative field modulus and phase distribution in stable vortex soliton 5 taken very close to the flat band is shown in Figure~\ref{fig2}(b).
	The solitons corresponding to dot 4 and dot 6 in Figure~\ref{fig2}(b) are more localized in comparison with these corresponding to dot 3 and dot 5.
	
	\begin{figure}[t]
		\centering
		\includegraphics[width=\columnwidth]{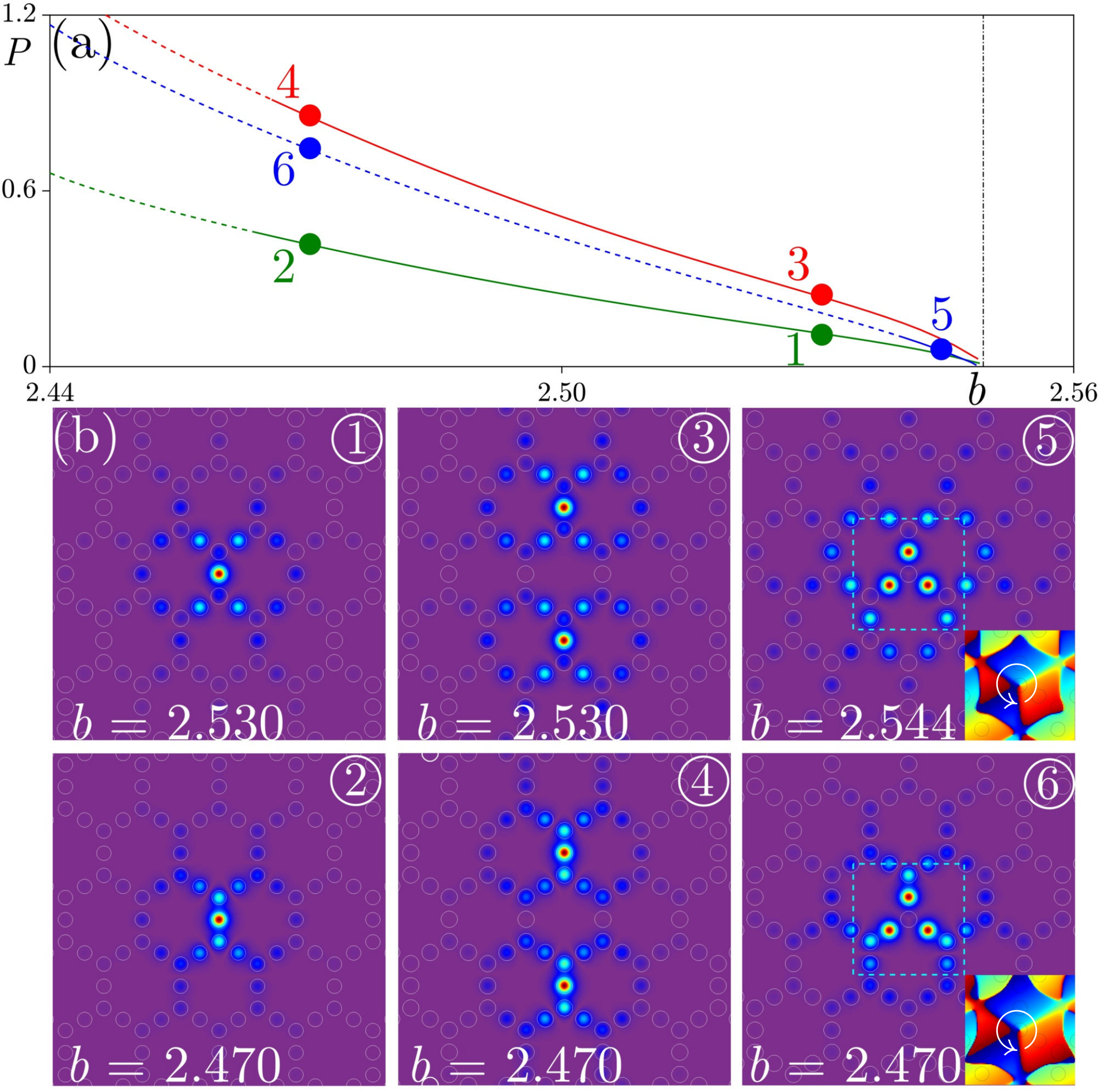}
		\caption{The families of solitons and examples of profiles in defocusing medium. Color coding of curves is the same as in Figure~\ref{fig2}. The vertical dashed line at ${b=2.5494}$ indicates the  location of the flat band, whose width cannot be discerned on the scale of the figure. The cut-off propagation constants of the red, green, and blue lines are $b=2.5488$, $b=2.549$, and $b=2.5486$, respectively.}
		\label{fig3}
	\end{figure}	
	
	The families of all such solitons share two common features. First, although gap solitons cannot bifurcate from linear Bloch modes (in the above mentioned sense) and exist above a certain power threshold, $P_{\rm th}$, this threshold becomes extremely small in flat-band system. 
	We successfully traced the families depicted in Figure~\ref{fig2}(a) practically up to the edge of the flat band. Further approaching $b_{\text{co}}$ became technically infeasible, even with very small increments of the propagation constant. Notice that even in such close proximity of the flat band, solitons remain well localized, in particular in comparison with solitons in lattices with dispersive bands that expand dramatically when their propagation constant approaches the edge of the band. Second, in a certain range of propagation constants near the flat band, the power $P$ becones a linear function of detuning $\delta b$, as discussed below (refer to Figure~\ref{fig4}).
	
	\subsection{Defocusing case}
	\label{sec:focus}
	
	From the physical point of view, a spatial soliton forms due to a delicate balance between diffraction and nonlinearity. Since in the leading order the diffraction is determined by the curvature of the allowed band, it is natural to expect that vanishing curvature of a flat band should result in similarity of properties of bright and dark solitons forming in the gap above and below such a band. This is fully confirmed by Figure~\ref{fig3} that shows soliton families [Figure~\ref{fig3}(a)] and shapes [Figure~\ref{fig3}(b)] of representative flat-band solitons in defocusing medium $(\sigma=-1)$, for the lattice depicted in Figure \ref{fig1}(c). Now solitons exist in the region ${\delta b<0}$, but except for this the obtained $P(b)$ dependencies in Figure \ref{fig3}(a) are remarkably similar to dependencies obtained for focusing case in Figure \ref{fig2}(a) (we use the same color coding for fundamental, dipole, and vortex solitons). Stability properties of soliton families are also similar, i.e. fundamental and dipole solitons in defocusing medium are stable in the largest part of the gap adjacent to the flat band, while vortex solitons are stable only a narrow region near the flat band. Just as in focusing medium analysis of $P(b)$ dependencies near the flat band reveals the presence of very small thresholds (comparable for solitons of different types) also in defocusing medium.
	
		It should be pointed out that while the dependencies $P(b)$ for flat-band solitons are qualitatively similar in focusing and defocusing media (except for the sign of the derivative $dP/db$  for the corresponding families), and the field modulus distributions for solitons shown in Figures \ref{fig2}(b) and \ref{fig3}(b) are also similar, the internal structure of solitons in focusing and defocusing media is different. This is particularly well visible from comparison of solutions 2 and 4 in Figures~\ref{fig2} and \ref{fig3}. One can see that in focusing medium the spots nearest to the central maximum of soliton are out-of-phase with central spot, while in defocusing medium they are in-phase with central spot. Therefore, the phrase structure in tails of flat-band solitons in focusing and defocusing media is clearly different. One can also see such differences from phase and field modulus distributions of vortex solitons in Figures \ref{fig2}(b) and \ref{fig3}(b).
	
	\section{Flat-band solitons: understanding the results}
	\label{sec:Wannier}
	
	\subsection{Wannier solitons}
	
	In order to explain the observed peculiarities of the flat-band solitons, we proceed with the asymptotic theory, corresponding to small but nonzero detuning: ${0<|\delta b|\ll b_{\rm co}}$. We are looking for nonlinear modes having propagation constants in the gap adjacent to the flat band and preserving strong localization (i.e., localization on the scale of a unit cell) even in the limit of negligible detuning ${|\delta b|\to 0}$. For construction of such solutions we must find a proper orthonormal basis consisting of localized sates.  
	
	Recalling that exact flat bands in discrete systems support compactons~\cite{yulin.ol.38.4880.2013} and allow in 2D case soliton families (exactly) bifurcating from the (exact) flat band~\cite{vicencio.pra.87.061803.2013}, and following the approach developed for 1D Wannier solitons~\cite{wang.pra.108.013307.2023}, here we employ 2D WFs
	\begin{align}
		\label{Wannier}
		w_{\nu}(a,\br) =  \frac{1}{|\Omega|}\int_{\rm BZ}e^{i\theta (a,\bk)}\varphi_{\nu \bk}(\br)d\bk
	\end{align}
	where ${|\Omega|=2\pi^2/(3\sqrt{3}d^2)}$ is the area of the BZ, $\nu$ is the flat band index, and $\theta (a,\bk)$ is an arbitrary phase, which as a function of $\bk$ has periodicity of the reciprocal lattice and depends on a parameter $a$ that we will discuss below. While the WFs are not uniquely defined, we do not impose the requirement on their best localization~\cite{marzari.rmp.84.1419.2012}. Furthermore, when a flat band touches other bands [see the lowest three bands in Figure~\ref{fig1}(e) and highest three bands in Figure~\ref{fig1}(f)] the exponentially localized WFs constructed using quasi-Bloch functions of composite bands~\cite{brouder.prl.98.046402.2007} appear inconvenient for our purposes because they camouflage the effect of diffraction suppression for Bloch modes of the flat band. Thus, the WF $w_{\nu}(a,\br)$ is considered localized in the central cell, i.e., in the cell with $\bn={\bf 0}$, although its specific decay with $|{\bm r}|$ is not specified.
	
	Using Equation~(\ref{Wannier}) it is straightforward to obtain 
	\begin{align}
		\label{Hw-aux}
		Hw_\nu(a,\br)=b_{\rm co}w_\nu(a,\br)+h(\br)\Delta_\nu ,
	\end{align}
	where 
	\begin{align}
		h(\br) &=\frac{1}{|\Omega|}\int_{\rm BZ}e^{i\theta (a,\bk)}\beta(\bk) \varphi_{\nu \bk}(\br)d\bk, \notag \\
		\beta(\bk) &=\frac{b_{\nu}(\bk)-b_{\rm co}}{\Delta_\nu}\in[-1,1].
	\end{align}
	Now, by analogy with the method adopted in \cite{wang.pra.108.013307.2023} for the 1D Wannier solitons, we perform the mixed Wannier-Bloch expansion of the searched solution of Equation~(\ref{NLS-stat}). Note that while technically expansions in 1D and 2D cases look similarly, an essential difference is that in the former case the expansion represents a soliton of a negligibly small amplitude, while in the last case such soliton does not exist~\cite{ilan.siam.8.1055.2010} below the threshold intensity. To this end, we introduce a formal small parameter ${0<\epsilon\ll 1}$, scaled propagation coordinates ${z_j=\epsilon z}$, which for ${j=0,1,...}$ are considered independent, so that ${\partial_z=\partial_{z_0}+\partial_{z_1}+\cdots}$, and look for a solution of Equation~(\ref{NLS}) in the form of the expansion
	\begin{align}
		\label{c-expan}
		\Psi= \sqrt{\epsilon}e^{ib_{\rm co} z_0}\left[A(a,z_1) w_\nu(a,\br) +\epsilon  \Psi_1 +\mathcal{O}(\epsilon^2)\right].
	\end{align}
	Here, $A(a,z_1)$ is a slowly varying amplitude of the soliton whose spatial profile in the leading order is described by the WF $w_\nu(A,\br)$ (it is assumed that the soliton is localized in the central cell) and
	\begin{align}
		\label{expan-psi}
		\Psi_1 = & \sum_{\bn\neq {\bm 0}} B_\bn(a,z_0)w_{\nu}(a,\br-\bv_\bn) + \notag \\
		& \sum_{\nu'\neq\nu} \int_{BZ}\,d\bk B_{\nu'} (\bk,z_0)\varphi_{\nu \bk}(\br)
	\end{align}
	is the first-order correction, $B_\bn(z_0)$ and $B_{\nu'} (\bk,z_0)$ are the amplitudes of the Wannier states of non-central cells (with $\bv_\bn\neq {\bf 0}$) of the flat band and of the Bloch states of other (non-flat) bands.
	
	Applying $H$ to both sides of the expansion Equation~(\ref{c-expan}), we obtain
	\begin{align}
		\label{H_psi}
		H\Psi=&
		\sqrt{\epsilon} e^{ib_{\rm co}  z_0} \big[ 
		b_{\rm co}  A(a,z_1) w_\nu(a,\br)+A(z_1) h(\br)\Delta_\nu  
		\nonumber \\
		&+\epsilon b_{\rm co} \sum_{\bn\neq {\bm 0}} B_\bn (a,z_0) w_\nu(a,\br-\bv_\bn) 
		\nonumber \\ 
		&+\epsilon\sum_{\nu'\neq\nu} \int_{BZ}d\bk B_{\nu'} (\bk,z_0)b_{\nu'}(\bk)\varphi_{\nu' \bk}(\br)\big]
	\end{align}
	Here, we neglected all terms of the order of $\epsilon^{5/2}$ and $\epsilon^{3/2}\Delta_\nu$, and used the fact that in the leading order Equation~(\ref{Hw-aux}) is valid for all WFs of the flat band (i.e., after the substitution ${\br\to\br-\bv_n}$). Thus, we have two small parameters: $\epsilon$ which is determined by the amplitude of the beam and $\Delta_\nu$, which is the system parameter characterising the flatness of the band. For the next steps we establish the hierarchy between them, by imposing the condition
	\begin{align}
		\label{DE}
		\Delta_\nu\ll \epsilon \ll 1.
	\end{align}
	On the one hand, this condition allows one to neglect the term $A(z_1) h(\br)\Delta_\nu$ in Equation~(\ref{H_psi}). On the other hand, the Equation~(\ref{DE}) means that the expansion is not applicable for beams with too small amplitudes. The width of the flat band in photonic lattices depicted in Figure~\ref{fig1} is ${\Delta_\nu<10^{-4}}$. Thus, for the validity of the expansion in Equation~(\ref{expan-psi}) the dimensionless soliton amplitude should satisfy ${|A(a,z_1)|\gtrsim\Delta_\nu^{1/2}\sim 10^{-2}}$. For solitons with smaller amplitudes one has to take into account the curvature of the flat band.  
	
	\begin{figure*}[htbp]
		\centering
		\includegraphics[width=1.5\columnwidth]{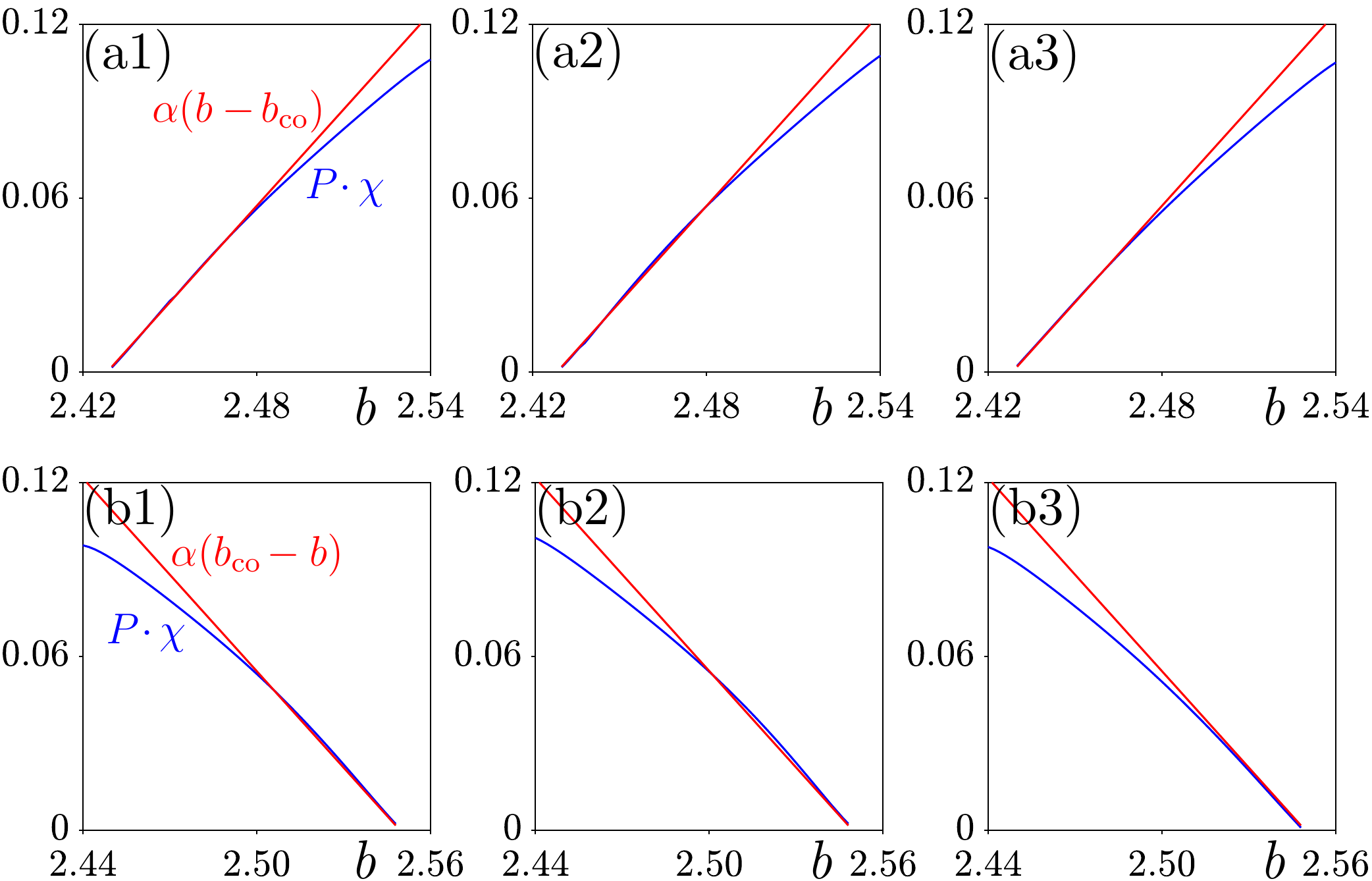}
		\caption{Comparison of flat-band soliton families obtained numerically (see blue curves corresponding to the numerically obtained soliton families shown in Figures~\ref{fig2} and \ref{fig3}) and theoretically (red lines, based on Equation~(\ref{Pb-exact})). Panels (a1-a3) correspond, respectively, to families of fundamental, dipole, and vortex solitons in focusing medium. Panels (b1-b3) correspond to fundamental, dipole, and vortex solitons, but in defocusing nonlinear medium.}
		\label{fig4}
	\end{figure*}
	
	Plugging the expansion from Equation~(\ref{H_psi}) into Equation~(\ref{NLS}), we obtain with the accuracy $\mathcal{O}(\epsilon)$
	\begin{align}
		\label{eq:a}
		& i\frac{\partial A(a,z_1)}{\partial z_1}w_\nu(a,\br) +   i\sum_{\bn\neq {\bm 0}}\frac{\partial B_\bn(a,z_0)}{\partial z_0} w_\nu(a,\br-\bv_\bn) 
		\nonumber \\ 
		& +i\sum_{\nu'\neq\nu} \int_{BZ}\,d\bk  \frac{\partial B_{\nu'} (\bk,z_0)}{\partial z_0} \varphi_{\nu' \bk} (\br) \nonumber \\
		= & \sum_{\nu'\neq\nu} \int_{BZ}\,d\bk B_{\nu'} (\bk,z_0)[b_{\rm co} - b_{\nu'}(\bk) ]\varphi_{\nu' \bk}(\br) \notag \\
		& -\sigma |A(a,z_1)|^2A(a,z_1) |w_\nu(a,\br)|^2w_\nu(a,\br).
	\end{align}
	Projecting over $w_\nu(a,\br)$ and using mutual orthogonality of the involved Wannier and Bloch functions, we arrive at the equation
	\begin{align}
		\label{eq:A}
		i \frac{\partial A}{\partial z_1} =-\sigma \chi(a) |A|^2A, 
	\end{align}
	where
	\begin{align}
		\chi(a)=\int_{\br\in\mathbb{R}^2} |w_\nu(a,\br)|^4 d\br
	\end{align}
	is the form-factor of the WF of the flat band. Thus, considering the solution of (\ref{eq:A}) in the form $A=ae^{i\sigma \chi(a) a^2z_1}$ where ${a>0}$ is an amplitude, what defines the physical meaning of this parameter introduced above in Equation~(\ref{Wannier}), the soliton constructed in accordance with the leading order of expansion (\ref{c-expan}) reads  
	\begin{align}
		\label{solut}
		\Psi=ae^{i b z}w_\nu(a,\br), \qquad b=b_{\rm co}+\sigma \chi |a|^2.
	\end{align}
	In this final expression we set the formal small parameter $\epsilon$ to one,  implying that the smallness is now ensured by the amplitude $a$. More specifically, we now require  ${|a|^2\ll b_{\rm co}/\chi}$ ensuring the relative smallness of the corrections to the propagation constant $b_{\rm co}$ in Equation~(\ref{solut}).  Since the profile of this soliton is determined by the WF, such a solution can be termed a  {\em Wannier soliton}~\cite{alfimov.pre.66.046608.2002, wang.pra.108.013307.2023}.   
	
	\subsection{Dependence of power on propagation constant}
	
	For the Wannier soliton (\ref{solut}) one obtains that the total power is $|a|^2$. This leads to the relation
	\begin{align}
		\label{Pb}
		P(a,b)\cdot \chi (a) \approx  \sigma \cdot \delta b. 
	\end{align}
	A peculiarity of this expression is that its right-hand side does not depend on the soliton amplitude: even though both power $P$ and form-factor $\chi$ depend on $a$, their product for the flat-band soliton is determined only by the detuning $\delta b$. Furthermore, the approximate expansion (\ref{c-expan}) does not imply bifurcation from a linear mode,  
	leaving freedom in the choice of a WF [the phase $\theta (a,\bk)$ introduced in (\ref{Wannier}) remains arbitrary] while the exact nonlinear solution for (\ref{NLS}) is well defined.  These factors not yet accounted in the expansion, strictly speaking, leave open exact applicability conditions for (\ref{Pb}). If however one conjecture that a flat-band soliton is approximated by some of WFs well enough, instead of searching for such specific WF, one can use the numerically found nonlinear solution instead of (\ref{solut}), interpreting the result (\ref{Pb}) as the relation
	\begin{align}
		\label{Pb-exact}
		P_s\cdot\chi_s \approx  \sigma \alpha \cdot \delta b,
	\end{align}
	where $P_s$ and $\chi_s$ are the power and the form-factor of the exact soliton, while $\alpha$ is the correction factor of order one. 

	We have checked the accuracy of formula (\ref{Pb-exact}) for all soliton families depicted in Figure~\ref{fig2} and~\ref{fig3}, and obtained a remarkable agreement illustrated in Figure~\ref{fig4}. In all cases the accurate linear fit of the families of fundamental, dipole, and vortex solitons was obtained with a universal factor ${\alpha=1.11}$.  
	
	\subsection{On excitation of extended modes}
	
	Now we briefly discuss the excitation of other modes in the system. Projecting Equation~(\ref{eq:a}) over $ {w_\nu(a,\br-\bv_\bn)}$ with ${\bv_\bn\neq {\bm 0}}$ we obtain
	\begin{align}
		&i\frac{\partial B_\bn(a,z_0)}{\partial z_0}= -\sigma  |A|^2A
		\notag \\	
		& \times
		\int w_\nu^* (a,\br-\bv_\bn)w_\nu(a,\br)|w_\nu(a,\br)|^2d\br.
	\end{align}
	This term gives a secular grows unless the nonlinear hopping is small enough, i.e., unless
	\begin{align}
		\label{hopping}
		\int w_\nu^* (a,\br-\bv_\bn)w_\nu(a,\br)|w_\nu(a,\br)|^2d\br \lesssim\epsilon \chi(a).
	\end{align}
	In that case, the energy transfer to WFs localized on non-central lattice sites with ${\bn\neq {\bm 0}}$ enabled by nonlinearity is described by the next order of the asymptotic expansion and we obtain ${\partial_{z_0} B_n=0}$. Thus, Equation~(\ref{hopping}) is the only condition on the localization of WFs, 
	that must be verified for the validity of the expansion in Equation~(\ref{c-expan}).
	
	Finally, projecting Equation~(\ref{eq:a}) on the Bloch state $\varphi_{\nu'k}$, we obtain 
	\begin{align}
		i\frac{\partial B_{\nu'} (\bk,z_0)}{\partial z_0}  
		= & B_{\nu'} (\bk,z_0)[b_{\rm co}- b_{\nu'}(\bk) ]  \notag \\
		& -\sigma \chi_{\nu'\bk}(a) |A(a,z_1)|^2A(a,z_1),
	\end{align}
	where
	\begin{align}
		{\chi_{\nu'\bk}(a)=\int\varphi_{\nu' \bk}^*(\br)|w_\nu(a,\br)|^2w_\nu(a,\br)d\br}.
	\end{align}
	Thus, non-flat bands do not give secularly growing terms, except for the Bloch state corresponding to the point where the flat band touches either lower or higher dispersive band (these are $\Gamma$ points in Figure~\ref{fig1}(e) and (f), correspondingly), where the equality ${b_{\rm co}=b_{\nu'}(\bk)} $ is verified. However, the weight (or measure) of the contribution of such modes in Equation~(\ref{expan-psi}) is negligible, and one can expect that they will not affect the leading  order solution Equation~(\ref{solut}).
	
	\begin{figure*}[htbp]
		\centering
		\includegraphics[width=\textwidth]{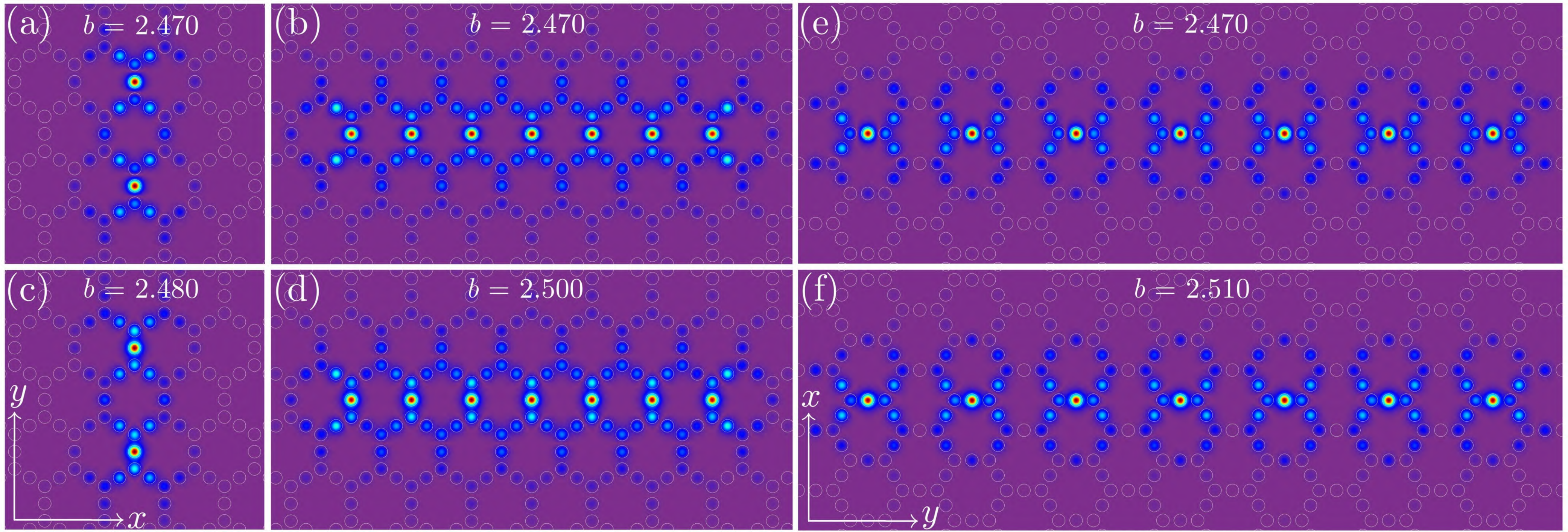}
		\caption{Examples of flat-band solitons of other types, including vertical 2-solitons (a,c), in-phase 7-solitons aligned in $x$ (b,d), and out-of-phase 7-solitons aligned in $y$ (e,f). Solutions (a,b) are obtained in focusing nonlinear medium, while solutions (c,d) are obtained in defocusing nonlinear medium. Solutions in (a,c) are shown within ${-15\le x,y\le15}$ window, solutions in (b,d) are shown within ${-30\le x\le30}$ and ${-15\le y\le15}$ window, while solutions in (e,f) are shown within ${-15\le x\le15}$ and ${-44\le y\le44}$ window.}
		\label{fig5}
	\end{figure*}
	
	\section{Flat-band multi-soliton solutions}
	\label{sec:multisol}
	
	Since flat-band (alias Wannier) solitons are strongly localized and can be stable in considerable part of the gap, it is natural to use them as building blocks for composition of multi-soliton solutions or even of soliton trains. Some solutions of this type are illustrated in Figure~\ref{fig5}. Solitons in Figure~\ref{fig5}(a) and (b) are obtained in focusing nonlinear medium, while solitons in Figure~\ref{fig5}(c) and (d) were found for defocusing nonlinearity. Using the in-phase superposition two fundamental solitons, i.e., ${\psi_1(y+\delta y) + \psi_2(y-\delta y)}$ (as opposed to out-of-phase combination yielding dipole solitons presented in Figure \ref{fig2} and \ref{fig3}) one obtains the 2-soliton (or even) solution depicted in Figure~\ref{fig5}(a). We also found various multi-soliton solutions aligned along the $x$-axis. Thus, in Figure~\ref{fig5}(b), we display a 7-soliton solution, which demonstrates that the flat-band solitons exist in abundant forms. Similar higher-order states exist and can be stable in defocusing medium, as shown in Figure~\ref{fig5}(c) and (d).
		While the 7-solitons in Figures~\ref{fig5}(b) and (d) are composed of in-phase states, out-of-phase higher-order multi-soliton solutions also exist. To obtain such multi-soliton solutions, we use an initial guess in the Newton method, combining several simple solitons with appropriately selected phase multipliers.  An example of such an out-of-phase solution, involving several elements aligned along the $y$-axis, is shown in Figures~\ref{fig5}(e) and (f). 
		The different phase structures of the soliton tails in focusing and defocusing media are also evident in these plots. The solutions presented in Figure \ref{fig5} can be considered as lattice soliton trains. It is noteworthy that lattice soliton trains have also been reported in non-flat-band systems~\cite{chen.prl.92.143902.2004}. However, the important difference from the states presented here is that intentional truncation or partial removal of the train in our case does not lead to significant shape variations of the remaining train upon propagation, whereas it can be quite destructive and lead to the decay of the train in non-flat-band systems. From an experimental point of view, the excitation of multi-soliton states involves creating the proper input, which can be achieved using interferometric techniques and spatial light modulators. Lattices of this type can be easily inscribed using fs-laser writing techniques in fused silica~\cite{rechtsman.nature.496.196.2013, kirsch.np.17.995.2021, arkhipova.sb.68.2017.2023, ren.light.12.194.2023, li.ap.4.024002.2022, li.light.12.262.2023, wang.light.13.130.2024}.
	
	\section{Conclusion}
	
	Summarizing, we have described properties of Wannier solitons that can be excited in two-dimensional continuous photonic lattices with a flat band in linear spectrum. Such flat-band solitons exist in both focusing and defocusing media. Their representative feature is that they remain well localized even in close proximity of the flat band. We have found several families of solitons, including fundamental, dipole, multi-soliton, and vortex ones. All these solutions can be stable in focusing or in defocusing medium despite the fact that they form in finite gap adjacent to the flat band. The analytical theory explaining the dependence of power of such solitons on propagation constant is developed for our continuous model. These results illustrate new nonlinear localization scenarios possible in real-world continuous flat-band periodic lattices with the translational symmetry (e.g., the Lieb lattice~\cite{vicencio.prl.114.245503.2015, mukherjee.prl.114.245504.2015, zhang.rrp.68.230.2016, xia.prl.121.263902.2018}, the Kagome lattice~\cite{boguslawski.apl.98.061111.2011, zhong.rip.12.996.2019, jiang.prb.99.125131.2019, zong.oe.24.8877.2016}, etc.).
	
	\section{Methods}
	
	The plane-wave expansion method is used to calculate the band structure shown in Figure~\ref{fig1}. The primitive vectors of the superhoneycomb lattice are  ${{\bm v}_1=[2\sqrt{3}d,0]^{\rm T}}$ and ${{\bm v}_2=[\sqrt{3}d,3d]^{\rm T}}$ ($\rm T$ stands for transpose)  with the angle between them ${\theta=\pi/3}$, as shown in Figure~\ref{fig1}(a). We then apply the following coordinate transformation in the direct and Fourier spaces:
		\begin{equation}
			\label{eq22}
			\br
			=\begin{bmatrix}
				1 & \cos \theta \\
				0 & \sin \theta \\
			\end{bmatrix}
			\br', 
			\qquad
			\bk'
			=\bk
			\begin{bmatrix}
				1 & \cos\theta \\
				0 & \sin\theta \\
			\end{bmatrix}.
		\end{equation}
		In new coordinate system the primitive lattice vectors become orthogonal and given by ${{\bm v}'_1=[2\sqrt{3}d,0]^{\rm T}}$ and ${{\bm v}'_2=[0,2\sqrt{3}d]^{\rm T}}$. In the new coordinate system, Equation~(\ref{linear}) can be rewritten as:
		\begin{equation}\label{eq23}
			b \varphi =  \frac{2}{3} \left(\frac{\partial^2}{\partial x'^2}+\frac{\partial^2}{\partial y'^2}-\frac{\partial^2}{{\partial x'}{\partial y'}} \right) \varphi + \mathcal{R}'(\br') \varphi,
		\end{equation}
		where ${\mathcal{R}'(\br')}$ is the superhoneycomb lattice in the frame $\br'$, i.e., after the coordinate transformation.
	
	We further expand $\varphi$ and $\mathcal{R}'$ into the Fourier series:
		\begin{equation}\label{eq24}
			\begin{split}
				\varphi = & e^{i\bk'\cdot \br'}\sum_{m,n}c_{m,n} e^{iK_m  x' + iK_n  y'},\\
				\mathcal{R}' = & \sum_{m,n}\rho_{m,n} e^{iK_m x' + iK_n y'},
			\end{split}
		\end{equation}
		where $c_{m,n}$ and $\rho_{m,n}$ are the Fourier coefficients with $m,n$ being integers, and ${K_{m}=m\pi/\sqrt{3}d}$. 
		Plugging expansions~(\ref{eq24}) into Equation~(\ref{eq23}), after straightforward algebra one obtains a system of linear equations: 
		\begin{equation*}
			\begin{split}
				& b c_{m, n} = \sum_{l,s} \rho_{l,s} c_{m-l, n-s} -\frac{2}{3} c_{m, n} \times \\
				& \left[(K_m + k'_x)^2+(K_n+k'_y)^2-(K_m + k'_x)(K_n+k'_y)\right] .
			\end{split}
		\end{equation*}
		Rewriting this equation in a matrix form and subsequently diagonalizing it, one obtains the eigenvalues $b(\bk')$ (i.e. the band structure) and the corresponding entries $c_{m,n}$ that allow one to construct the eigenmodes $u$ of the array using the expressions~(\ref{eq24}).
	
	
	\begin{funding}
		This work was supported by the Natural Science Basic Research Program of Shaanxi Province (2024JC-JCQN-06), the National Natural Science Foundation of China (12074308), the research project FFUU-2024-0003 of the Institute of Spectroscopy of the Russian Academy of Sciences, and the Fundamental Research Funds for the Central Universities (xzy022023059). The work of V.V.K. was supported by the Portuguese Foundation for Science and Technology (FCT) under Contracts UIDB/00618/2020 (doi: 10.54499/UIDB/00618/2020) and PTDC/FIS-OUT/3882/2020 (doi: 10.54499/PTDC/FIS-OUT/3882/2020).
	\end{funding}
	
	\begin{authorcontributions}
		All the authors have accepted responsibility for the entire content of this submitted
		manuscript and approved submission.
	\end{authorcontributions}
	
	\begin{conflictofinterest}
		The authors state no conflict of interest.
	\end{conflictofinterest}
	
	\begin{dataavailabilitystatement}
		The datasets generated during and/or analyzed during the current study are available from the corresponding author on reasonable request.
	\end{dataavailabilitystatement}
	


\begin{thebibliography}{10}
		\providecommand{\url}[1]{#1}
		\csname url@samestyle\endcsname
		\providecommand{\newblock}{\relax}
		\providecommand{\bibinfo}[2]{#2}
		\providecommand{\BIBentrySTDinterwordspacing}{\spaceskip=0pt\relax}
		\providecommand{\BIBentryALTinterwordstretchfactor}{4}
		\providecommand{\BIBentryALTinterwordspacing}{\spaceskip=\fontdimen2\font plus
			\BIBentryALTinterwordstretchfactor\fontdimen3\font minus
			\fontdimen4\font\relax}
		\providecommand{\BIBforeignlanguage}[2]{{%
				\expandafter\ifx\csname l@#1\endcsname\relax
				\typeout{** WARNING: IEEEtran.bst: No hyphenation pattern has been}%
				\typeout{** loaded for the language `#1'. Using the pattern for}%
				\typeout{** the default language instead.}%
				\else
				\language=\csname l@#1\endcsname
				\fi
				#2}}
		\providecommand{\BIBdecl}{\relax}
		\BIBdecl
		
		\bibitem{leykam.aplp.3.070901.2018}
		\BIBentryALTinterwordspacing
		D.~Leykam and S.~Flach, ``Perspective: Photonic flatbands,'' \emph{APL
			Photon.}, vol.~3, no.~7, p. 070901, 06 2018.
		
		\bibitem{leykam.aipx.3.1473052.2018}
		D.~Leykam, A.~Andreanov, and S.~Flach, ``Artificial flat band systems: from
		lattice models to experiments,'' \emph{Adv. Phys. X}, vol.~3, no.~1, p.
		1473052, 2018.
		
		\bibitem{tang.nano.9.1161.2020}
		L.~Tang, D.~Song, S.~Xia, S.~Xia, J.~Ma, W.~Yan, Y.~Hu, J.~Xu, D.~Leykam, and
		Z.~Chen, ``Photonic flat-band lattices and unconventional light
		localization,'' \emph{Nanophoton.}, vol.~9, no.~5, pp. 1161--1176, 2020.
		
		\bibitem{vicencio.aipx.6.1878057.2021}
		R.~A.~V. Poblete, ``Photonic flat band dynamics,'' \emph{Adv. Phys. X}, vol.~6,
		no.~1, p. 1878057, 2021.
		
		\bibitem{chekelsky.nrm.2024}
		\BIBentryALTinterwordspacing
		J.~G. Checkelsky, B.~A. Bernevig, P.~Coleman, Q.~Si, and S.~Paschen, ``Flat
		bands, strange metals and the {Kondo} effect,'' \emph{Nat. Rev. Mater.}, Feb.
		2024. 
		
		\bibitem{morales.pra.94.043831.2016}
		\BIBentryALTinterwordspacing
		L.~Morales-Inostroza and R.~A. Vicencio, ``Simple method to construct flat-band
		lattices,'' \emph{Phys. Rev. A}, vol.~94, p. 043831, Oct 2016.
		
		\bibitem{vicencio.prl.114.245503.2015}
		\BIBentryALTinterwordspacing
		R.~A. Vicencio, C.~Cantillano, L.~Morales-Inostroza, B.~Real,
		C.~Mej\'{i}a-Cort\'es, S.~Weimann, A.~Szameit, and M.~I. Molina,
		``Observation of localized states in {L}ieb photonic lattices,'' \emph{Phys.
			Rev. Lett.}, vol. 114, p. 245503, Jun 2015.
		
		\bibitem{mukherjee.prl.114.245504.2015}
		\BIBentryALTinterwordspacing
		S.~Mukherjee, A.~Spracklen, D.~Choudhury, N.~Goldman, P.~\"Ohberg,
		E.~Andersson, and R.~R. Thomson, ``Observation of a localized flat-band state
		in a photonic {L}ieb lattice,'' \emph{Phys. Rev. Lett.}, vol. 114, p. 245504,
		Jun 2015.
		
		\bibitem{zhang.aop.382.160.2017}
		\BIBentryALTinterwordspacing
		D.~Zhang, Y.~Zhang, H.~Zhong, C.~Li, Z.~Zhang, Y.~Zhang, and M.~R. Beli\'c,
		``New edge-centered photonic square lattices with flat bands,'' \emph{Ann.
			Phys.}, vol. 382, pp. 160--169, 2017.
		
		\bibitem{hanafi.apl.7.111301.2022}
		\BIBentryALTinterwordspacing
		H.~Hanafi, P.~Menz, A.~McWilliam, J.~Imbrock, and C.~Denz, ``Localized dynamics
		arising from multiple flat bands in a decorated photonic {Lieb} lattice,''
		\emph{APL Photon.}, vol.~7, no.~11, p. 111301, 11 2022.
		
		\bibitem{xia.ol.41.1435.2016}
		\BIBentryALTinterwordspacing
		S.~Xia, Y.~Hu, D.~Song, Y.~Zong, L.~Tang, and Z.~Chen, ``Demonstration of
		flat-band image transmission in optically induced {L}ieb photonic lattices,''
		\emph{Opt. Lett.}, vol.~41, pp. 1435--1438, Apr 2016.
		
		\bibitem{zong.oe.24.8877.2016}
		\BIBentryALTinterwordspacing
		Y.~Zong, S.~Xia, L.~Tang, D.~Song, Y.~Hu, Y.~Pei, J.~Su, Y.~Li, and Z.~Chen,
		``Observation of localized flat-band states in kagome photonic lattices,''
		\emph{Opt. Express}, vol.~24, pp. 8877--8885, Apr 2016.
		
		\bibitem{travkin.apl.111.011104.2017}
		E.~Travkin, F.~Diebel, and C.~Denz, ``Compact flat band states in optically
		induced flatland photonic lattices,'' \emph{Appl. Phys. Lett.}, vol. 111,
		no.~1, p. 011104, 07 2017.
		
		\bibitem{xia.prl.121.263902.2018}
		\BIBentryALTinterwordspacing
		S.~Xia, A.~Ramachandran, S.~Xia, D.~Li, X.~Liu, L.~Tang, Y.~Hu, D.~Song, J.~Xu,
		D.~Leykam, S.~Flach, and Z.~Chen, ``Unconventional flatband line states in
		photonic {Lieb} lattices,'' \emph{Phys. Rev. Lett.}, vol. 121, p. 263902, Dec
		2018. 
		
		\bibitem{ma.prl.124.183901.2020}
		\BIBentryALTinterwordspacing
		J.~Ma, J.-W. Rhim, L.~Tang, S.~Xia, H.~Wang, X.~Zheng, S.~Xia, D.~Song, Y.~Hu,
		Y.~Li, B.-J. Yang, D.~Leykam, and Z.~Chen, ``Direct observation of flatband
		loop states arising from nontrivial real-space topology,'' \emph{Phys. Rev.
			Lett.}, vol. 124, p. 183901, May 2020.
		
		\bibitem{yan.aom.8.1902174.2020}
		\BIBentryALTinterwordspacing
		W.~Yan, H.~Zhong, D.~Song, Y.~Zhang, S.~Xia, L.~Tang, D.~Leykam, and Z.~Chen,
		``Flatband line states in photonic super-honeycomb lattices,'' \emph{Adv.
			Opt. Mater.}, vol.~8, no.~11, p. 1902174, 2020.
		
		\bibitem{song.lpr.17.2200315.2023}
		\BIBentryALTinterwordspacing
		L.~Song, Y.~Xie, S.~Xia, L.~Tang, D.~Song, J.-W. Rhim, and Z.~Chen,
		``Topological flatband loop states in fractal-like photonic lattices,''
		\emph{Laser Photon. Rev.}, vol.~17, no.~8, p. 2200315, 2023. 
		
		\bibitem{yang.nc.15.1484.2024}
		\BIBentryALTinterwordspacing
		J.~Yang, Y.~Li, Y.~Yang, X.~Xie, Z.~Zhang, J.~Yuan, H.~Cai, D.-W. Wang, and
		F.~Gao, ``Realization of all-band-flat photonic lattices,'' \emph{Nat.
			Commun.}, vol.~15, no.~1, p. 1484, 2024.
		
		\bibitem{vicencio.pra.87.061803.2013}
		\BIBentryALTinterwordspacing
		R.~A. Vicencio and M.~Johansson, ``Discrete flat-band solitons in the kagome
		lattice,'' \emph{Phys. Rev. A}, vol.~87, p. 061803, Jun 2013. 
		
		\bibitem{yulin.ol.38.4880.2013}
		\BIBentryALTinterwordspacing
		A.~V. Yulin and V.~V. Konotop, ``Conservative and {PT}-symmetric compactons in
		waveguide networks,'' \emph{Opt. Lett.}, vol.~38, no.~22, pp. 4880--4883, Nov
		2013.
		
		\bibitem{johansson.pre.92.032912.2015}
		\BIBentryALTinterwordspacing
		M.~Johansson, U.~Naether, and R.~A. Vicencio, ``Compactification tuning for
		nonlinear localized modes in sawtooth lattices,'' \emph{Phys. Rev. E},
		vol.~92, p. 032912, Sep 2015.
		
		\bibitem{gligoric.prb.94.144302.2016}
		\BIBentryALTinterwordspacing
		G.~Gligori\ifmmode~\acute{c}\else \'{c}\fi{}, A.~Maluckov,
		L.~Had\ifmmode~\check{z}\else \v{z}\fi{}ievski, S.~Flach, and B.~A. Malomed,
		``Nonlinear localized flat-band modes with spin-orbit coupling,'' \emph{Phys.
			Rev. B}, vol.~94, p. 144302, Oct 2016.
		
		\bibitem{zegadlo.pre.96.012204.2017}
		\BIBentryALTinterwordspacing
		K.~Zegadlo, N.~Dror, N.~Viet~Hung, M.~Trippenbach, and B.~A. Malomed, ``Single
		and double linear and nonlinear flatband chains: Spectra and modes,''
		\emph{Phys. Rev. E}, vol.~96, p. 012204, Jul 2017.
		
		\bibitem{baboux.prl.116.066402.2016}
		\BIBentryALTinterwordspacing
		F.~Baboux, L.~Ge, T.~Jacqmin, M.~Biondi, E.~Galopin, A.~Lema\^{\i}tre,
		L.~Le~Gratiet, I.~Sagnes, S.~Schmidt, H.~E. T\"ureci, A.~Amo, and J.~Bloch,
		``Bosonic condensation and disorder-induced localization in a flat band,''
		\emph{Phys. Rev. Lett.}, vol. 116, p. 066402, Feb 2016.
		
		\bibitem{goblot.prl.123.113901.2019}
		\BIBentryALTinterwordspacing
		V.~Goblot, B.~Rauer, F.~Vicentini, A.~Le~Boit\'e, E.~Galopin, A.~Lema\^{\i}tre,
		L.~Le~Gratiet, A.~Harouri, I.~Sagnes, S.~Ravets, C.~Ciuti, A.~Amo, and
		J.~Bloch, ``Nonlinear polariton fluids in a flatband reveal discrete gap
		solitons,'' \emph{Phys. Rev. Lett.}, vol. 123, p. 113901, Sep 2019.
		
		\bibitem{belicev.pra.96.063838.2017}
		\BIBentryALTinterwordspacing
		P.~P. Beli\ifmmode~\check{c}\else \v{c}\fi{}ev,
		G.~Gligori\ifmmode~\acute{c}\else \'{c}\fi{}, A.~Maluckov,
		M.~Stepi\ifmmode~\acute{c}\else \'{c}\fi{}, and M.~Johansson, ``Localized gap
		modes in nonlinear dimerized {Lieb} lattices,'' \emph{Phys. Rev. A}, vol.~96,
		p. 063838, Dec 2017.
		
		\bibitem{lazarides.prb.96.054305.2017}
		\BIBentryALTinterwordspacing
		N.~Lazarides and G.~P. Tsironis, ``{SQUID} metamaterials on a {Lieb} lattice:
		From flat-band to nonlinear localization,'' \emph{Phys. Rev. B}, vol.~96, p.
		054305, Aug 2017.
		
		\bibitem{real.pra.98.053845.2018}
		\BIBentryALTinterwordspacing
		B.~Real and R.~A. Vicencio, ``Controlled mobility of compact discrete solitons
		in nonlinear {Lieb} photonic lattices,'' \emph{Phys. Rev. A}, vol.~98, p.
		053845, Nov 2018.
		
		\bibitem{ali.pra.103.013517.2021}
		\BIBentryALTinterwordspacing
		A.~K.~S. Ali, A.~I. Maimistov, K.~Porsezian, A.~Govindarajan, and
		M.~Lakshmanan, ``Modulational instability in a non-kerr photonic {Lieb}
		lattice with metamaterials,'' \emph{Phys. Rev. A}, vol. 103, p. 013517, Jan
		2021.
		
		\bibitem{stojanovic.pra.102.023532.2020}
		\BIBentryALTinterwordspacing
		M.~G. Stojanovi\ifmmode~\acute{c}\else \'{c}\fi{}, M.~S. a.~c.
		Krasi\ifmmode~\acute{c}\else \'{c}\fi{}, A.~Maluckov, M.~Johansson, I.~A.
		Salinas, R.~A. Vicencio, and M.~Stepi\ifmmode~\acute{c}\else \'{c}\fi{},
		``Localized modes in linear and nonlinear octagonal-diamond lattices with two
		flat bands,'' \emph{Phys. Rev. A}, vol. 102, p. 023532, Aug 2020.
		
		\bibitem{mihalache.rrp.76.402.2024}
		D.~Mihalache, ``Localized structures in optical media and {Bose-Einstein}
		condensates: An overview of recent theoretical and experimental results,''
		\emph{Rom. Rep. Phys.}, vol.~76, p. 402, 2024.
		
		\bibitem{wang.pra.108.013307.2023}
		\BIBentryALTinterwordspacing
		C.~Wang, Y.~Zhang, and V.~V. Konotop, ``Wannier solitons in spin-orbit-coupled
		{Bose-Einstein} condensates in optical lattices with a flat band,''
		\emph{Phys. Rev. A}, vol. 108, p. 013307, Jul 2023.
		
		\bibitem{wang.nature.577.42.2020}
		\BIBentryALTinterwordspacing
		P.~Wang, Y.~Zheng, X.~Chen, C.~Huang, Y.~V. Kartashov, L.~Torner, V.~V.
		Konotop, and F.~Ye, ``Localization and delocalization of light in photonic
		moir\'e lattices,'' \emph{Nature}, vol. 577, no. 7788, pp. 42--46, Jan. 2020.
		
		\bibitem{fu.np.14.663.2020}
		\BIBentryALTinterwordspacing
		Q.~Fu, P.~Wang, C.~Huang, Y.~V. Kartashov, L.~Torner, V.~V. Konotop, and F.~Ye,
		``Optical soliton formation controlled by angle twisting in photonic moir\'e
		lattices,'' \emph{Nat. Photon.}, vol.~14, no.~11, pp. 663--668, 2020.
		
		\bibitem{marzari.rmp.84.1419.2012}
		\BIBentryALTinterwordspacing
		N.~Marzari, A.~A. Mostofi, J.~R. Yates, I.~Souza, and D.~Vanderbilt,
		``Maximally localized {Wannier} functions: Theory and applications,''
		\emph{Rev. Mod. Phys.}, vol.~84, pp. 1419--1475, Oct 2012.
		
		\bibitem{zhong.adp.529.1600258.2017}
		\BIBentryALTinterwordspacing
		H.~Zhong, Y.~Q. Zhang, Y.~Zhu, D.~Zhang, C.~B. Li, Y.~P. Zhang, F.~L. Li, M.~R.
		Beli\'c, and M.~Xiao, ``Transport properties in the photonic super-honeycomb
		lattice -- a hybrid fermionic and bosonic system,'' \emph{Ann. Phys.
			(Berlin)}, vol. 529, no.~3, p. 1600258, 2017.
		
		\bibitem{lan.prb.85.155451.2012}
		\BIBentryALTinterwordspacing
		Z.~Lan, N.~Goldman, and P.~\"Ohberg, ``Coexistence of spin-$\frac{1}{2}$ and
		spin-1 {Dirac-Weyl} fermions in the edge-centered honeycomb lattice,''
		\emph{Phys. Rev. B}, vol.~85, p. 155451, Apr 2012.
		
		\bibitem{rechtsman.nature.496.196.2013}
		\BIBentryALTinterwordspacing
		M.~C. Rechtsman, J.~M. Zeuner, Y.~Plotnik, Y.~Lumer, D.~Podolsky, F.~Dreisow,
		S.~Nolte, M.~Segev, and A.~Szameit, ``Photonic {F}loquet topological
		insulators,'' \emph{Nature}, vol. 496, pp. 196--200, 2013.
		
		\bibitem{kirsch.np.17.995.2021}
		\BIBentryALTinterwordspacing
		M.~S. Kirsch, Y.~Zhang, M.~Kremer, L.~J. Maczewsky, S.~K. Ivanov, Y.~V.
		Kartashov, L.~Torner, D.~Bauer, A.~Szameit, and M.~Heinrich, ``Nonlinear
		second-order photonic topological insulators,'' \emph{Nat. Phys.}, vol.~17,
		no.~9, pp. 995--1000, Sep. 2021.
		
		\bibitem{ren.light.12.194.2023}
		\BIBentryALTinterwordspacing
		B.~Ren, A.~A. Arkhipova, Y.~Zhang, Y.~V. Kartashov, H.~Wang, S.~A.
		Zhuravitskii, N.~N. Skryabin, I.~V. Dyakonov, A.~A. Kalinkin, S.~P. Kulik,
		V.~O. Kompanets, S.~V. Chekalin, and V.~N. Zadkov, ``Observation of nonlinear
		disclination states,'' \emph{Light Sci. Appl.}, vol.~12, no.~1, p. 194, Aug
		2023.
		
		\bibitem{arkhipova.sb.68.2017.2023}
		\BIBentryALTinterwordspacing
		A.~A. Arkhipova, Y.~Zhang, Y.~V. Kartashov, S.~A. Zhuravitskii, N.~N. Skryabin,
		I.~V. Dyakonov, A.~A. Kalinkin, S.~P. Kulik, V.~O. Kompanets, S.~V. Chekalin,
		and V.~N. Zadkov, ``Observation of $\pi$ solitons in oscillating waveguide
		arrays,'' \emph{Sci. Bull.}, vol.~68, no.~18, pp. 2017--2024, 2023. 
		
		\bibitem{zhong.ap.3.056001.2021}
		\BIBentryALTinterwordspacing
		H.~Zhong, S.~Xia, Y.~Zhang, Y.~Li, D.~Song, C.~Liu, and Z.~Chen, ``{Nonlinear
			topological valley {Hall} edge states arising from type-{II} {Dirac}
			cones},'' \emph{Adv. Photon.}, vol.~3, no.~5, p. 056001, 2021. 
		
		\bibitem{zhang.elight.3.5.2023}
		\BIBentryALTinterwordspacing
		Y.~Zhang, D.~Bongiovanni, Z.~Wang, X.~Wang, S.~Xia, Z.~Hu, D.~Song, D.~Juki\'c,
		J.~Xu, R.~Morandotti, H.~Buljan, and Z.~Chen, ``Realization of photonic
		$p$-orbital higher-order topological insulators,'' \emph{eLight}, vol.~3,
		no.~1, p.~5, Mar. 2023.
		
		\bibitem{wang.np.18.224.2024}
		\BIBentryALTinterwordspacing
		P.~Wang, Q.~Fu, V.~V. Konotop, Y.~V. Kartashov, and F.~Ye, ``Observation of
		localization of light in linear photonic quasicrystals with diverse
		rotational symmetries,'' vol.~18, no.~3, pp. 224--229.
		
		\bibitem{tang.rrp.74.504.2022}
		Q.~Tang, B.~Ren, M.~R. Beli\'c, Y.~Zhang, and Y.~Li, ``Valley {Hall} edge
		solitons in the kagome photonic lattice,'' \emph{Rom. Rep. Phys.}, vol.~74,
		p. 504, 2022.
		
		\bibitem{ilan.siam.8.1055.2010}
		\BIBentryALTinterwordspacing
		B.~Ilan and M.~I. Weinstein, ``Band-edge solitons, nonlinear
		{Schr\"odinger/Gross–Pitaevskii} equations, and effective media,''
		\emph{Multiscale Modeling \& Simulation}, vol.~8, no.~4, pp. 1055--1101,
		2010.
		
		\bibitem{shi.pre.75.056602.2007}
		\BIBentryALTinterwordspacing
		Z.~Shi and J.~Yang, ``Solitary waves bifurcated from {Bloch}-band edges in
		two-dimensional periodic media,'' \emph{Phys. Rev. E}, vol.~75, p. 056602,
		May 2007.
		
		\bibitem{shi.pra.78.063812.2008}
		\BIBentryALTinterwordspacing
		Z.~Shi, J.~Wang, Z.~Chen, and J.~Yang, ``Linear instability of two-dimensional
		low-amplitude gap solitons near band edges in periodic media,'' \emph{Phys.
			Rev. A}, vol.~78, p. 063812, Dec 2008.
		
		\bibitem{yang.book.2010}
		J.~Yang, \emph{Nonlinear Waves in Integrable and Non-Integrable Systems}.\hskip
		1em plus 0.5em minus 0.4em\relax Philadelphia: SIAM, 2010.
		
		\bibitem{brouder.prl.98.046402.2007}
		\BIBentryALTinterwordspacing
		C.~Brouder, G.~Panati, M.~Calandra, C.~Mourougane, and N.~Marzari,
		``Exponential localization of {Wannier} functions in insulators,''
		\emph{Phys. Rev. Lett.}, vol.~98, p. 046402, Jan 2007.
		
		\bibitem{alfimov.pre.66.046608.2002}
		\BIBentryALTinterwordspacing
		G.~L. Alfimov, P.~G. Kevrekidis, V.~V. Konotop, and M.~Salerno, ``Wannier
		functions analysis of the nonlinear {Schr\"odinger} equation with a periodic
		potential,'' \emph{Phys. Rev. E}, vol.~66, p. 046608, Oct 2002.
		
		\bibitem{chen.prl.92.143902.2004}
		\BIBentryALTinterwordspacing
		Z.~Chen, H.~Martin, E.~D. Eugenieva, J.~Xu, and A.~Bezryadina, ``Anisotropic
		enhancement of discrete diffraction and formation of two-dimensional
		discrete-soliton trains,'' \emph{Phys. Rev. Lett.}, vol.~92, p. 143902, Apr
		2004.
		
		\bibitem{li.ap.4.024002.2022}
		\BIBentryALTinterwordspacing
		L.~Li, W.~Kong, and F.~Chen, ``Femtosecond laser-inscribed optical waveguides
		in dielectric crystals: a concise review and recent advances,'' \emph{Adv.
			Photon.}, vol.~4, no.~11, p. 024002, 2 2022. 
		
		\bibitem{li.light.12.262.2023}
		\BIBentryALTinterwordspacing
		M.~Li, C.~Li, L.~Yan, Q.~Li, Q.~Gong, and Y.~Li, ``Fractal photonic anomalous
		{Floquet} topological insulators to generate multiple quantum chiral edge
		states,'' \emph{Light Sci. Appl.}, vol.~12, no.~1, p. 262, Nov. 2023.
		
		\bibitem{wang.light.13.130.2024}
		\BIBentryALTinterwordspacing
		Y.~Wang, L.~Zhong, K.~Y. Lau, X.~Han, Y.~Yang, J.~Hu, S.~Firstov, Z.~Chen,
		Z.~Ma, L.~Tong, K.~S. Chiang, D.~Tan, and J.~Qiu, ``Precise mode control of
		laser-written waveguides for broadband, low-dispersion {3D} integrated
		optics,'' \emph{Light Sci. Appl.}, vol.~13, no.~1, p. 130, Jun. 2024.
		
		\bibitem{zhang.rrp.68.230.2016}
		Y.~Q. Zhang, X.~Liu, M.~Beli\'c, W.~P. Zhong, C.~B. Li, H.~X. Chen, and Y.~P.
		Zhang, ``Dispersion relations of strained and complex lieb lattices based on
		the tight-binding method,'' \emph{Rom. Rep. Phys.}, vol.~68, pp. 230--240,
		2016.
		
		\bibitem{boguslawski.apl.98.061111.2011}
		\BIBentryALTinterwordspacing
		M.~Boguslawski, P.~Rose, and C.~Denz, ``Nondiffracting kagome lattice,''
		\emph{Appl. Phys. Lett.}, vol.~98, p. 061111, 2011.
		
		\bibitem{zhong.rip.12.996.2019}
		H.~Zhong, R.~Wang, F.~Ye, J.~Zhang, L.~Zhang, Y.~P. Zhang, M.~R. Beli\'c, and
		Y.~Q. Zhang, ``Topological insulator properties of photonic kagome helical
		waveguide arrays,'' \emph{Results Phys.}, vol.~12, pp. 996--1001, 2019.
		
		\bibitem{jiang.prb.99.125131.2019}
		\BIBentryALTinterwordspacing
		W.~Jiang, M.~Kang, H.~Huang, H.~Xu, T.~Low, and F.~Liu, ``Topological band
		evolution between {Lieb} and kagome lattices,'' \emph{Phys. Rev. B}, vol.~99,
		p. 125131, Mar 2019.
		
	\end{thebibliography}

\end{document}